\documentclass[aps,pra,reprint,groupedaddress,superscriptaddress,longbibliography]{revtex4-1}
\usepackage{mathrsfs}
\usepackage{blindtext}
\usepackage{soul}
\usepackage{amssymb,amsbsy,amsmath,latexsym,dsfont,array,layout,graphicx,mathrsfs,color,ulem,bm}

\begin{document}

\title{Quantum information dynamics in a high-dimensional parity-time-symmetric system}
\author{Zhihao Bian}
\affiliation{Beijing Computational Science Research Center, Beijing 100084, China}
\affiliation{School of Science, Jiangnan University, Wuxi 214122, China}
\author{Lei Xiao}
\affiliation{Beijing Computational Science Research Center, Beijing 100084, China}
\author{Kunkun Wang}
\affiliation{Beijing Computational Science Research Center, Beijing 100084, China}
\author{Franck Assogba Onanga}
\affiliation{Department of Physics, Indiana University Purdue University Indianapolis (IUPUI), Indianapolis, Indiana 46202, USA}
\author{Frantisek Ruzicka}
\affiliation{Department of Physics, Indiana University Purdue University Indianapolis (IUPUI), Indianapolis, Indiana 46202, USA}
\affiliation{Institute of Nuclear Physics, Czech Academy of Sciences, Rez 250 68, Czech Republic}
\author{Wei Yi}\email{wyiz@ustc.edu.cn}
\affiliation{CAS Key Laboratory of Quantum Information, University of Science and Technology of China, Hefei 230026, China}
\affiliation{CAS Center For Excellence in Quantum Information and Quantum Physics, Hefei 230026, China}
\author{Yogesh N. Joglekar}\email{yojoglek@iupui.edu}
\affiliation{Department of Physics, Indiana University Purdue University Indianapolis (IUPUI), Indianapolis, Indiana 46202, USA}
\author{Peng Xue}\email{gnep.eux@gmail.com}
\affiliation{Beijing Computational Science Research Center, Beijing 100084, China}

\begin{abstract}
Non-Hermitian systems with parity-time ($\mathcal{PT}$) symmetry give rise to exceptional points (EPs) with exceptional properties that arise due to the coalescence of eigenvectors. Such systems have been extensively explored in the classical domain, where second or higher order EPs have been proposed or realized. In contrast, quantum information studies of $\mathcal{PT}$-symmetric systems have been confined to systems with a two-dimensional Hilbert space. Here by using a single-photon interferometry setup, we simulate quantum dynamics of a four-dimensional $\mathcal{PT}$-symmetric system across a fourth-order exceptional point. By tracking the coherent, non-unitary evolution of the density matrix of the system in $\mathcal{PT}$-symmetry unbroken and broken regions, we observe the entropy dynamics for both the entire system, and the gain and loss subsystems. Our setup is scalable to the higher-dimensional $\mathcal{PT}$-symmetric systems, and our results point towards the rich dynamics and critical properties.
\end{abstract}

\maketitle

%\label{sec:intro}  % \label{} allows reference to this section
\section{Introduction}

A fundamental postulate of quantum theory is that the Hamiltonian of an isolated system is Hermitian. This Hermiticity ensures real eigenvalues and a coherent, unitary time evolution for the system. This conventional wisdom was upended two decades ago by Carl Bender and co-workers, who showed that a non-Hermitian Hamiltonian with parity-time ($\mathcal{PT}$) symmetry can exhibit entirely real spectra~\cite{CDH02,A02,A10,CHS+15}. Over time, it has become clear that non-Hermitian Hamiltonians with $\mathcal{PT}$ symmetry can provide an effective description for systems
with balanced, spatially separated gain and loss~\cite{JTSV13}. This concept has been extensively, and fruitfully, explored in classical (wave) systems where the number of energy quanta is much larger than one~\cite{CKR+10,alois2012,feng2014,BSF+14,CJH+14,BLD+14,WKP+16,AYF17,LRL17,HAS+17}. A $\mathcal{PT}$-symmetric system is described by an effective, non-Hermitian Hamiltonian $H_\mathcal{PT}$ that is invariant under the combined parity and time-reversal operation~\cite{CS98}. As the gain-loss strength is increased, the spectrum of $H_\mathcal{PT}$ changes from real into complex conjugate pairs, and the corresponding eigenvectors cease to be eigenvectors of the $\mathcal{PT}$ operator. This $\mathcal{PT}$-symmetry-breaking transition occurs at an EP of order $n$ (EP$n$), where $n$ eigenvalues, as well as their corresponding eigenvectors, coalesce~\cite{kato56,W12,ORN+19}. The $\mathcal{PT}$ transition and the non-unitary time evolution generated by $H_\mathcal{PT}$ have been observed in classical systems with EP$2$~\cite{CKR+10,alois2012,feng2014,BSF+14,CJH+14,BLD+14,WKP+16,AYF17,LRL17,POR+14,ZPO+18,MMC+19,MMC+19,MMC+20,JMC+19,HLL+19,MOB+15}, EP$3$~\cite{HAS+17}, and higher order EPs~\cite{XLKA19,JOL+17}.

Due to the quantum limit on noise in linear (gain) amplifiers~\cite{Caves1982}, creating a photonic system with  balanced gain and loss in the quantum domain is not possible~\cite{Scheel2018}. However, the EP degeneracies also occur in dissipative systems with mode-selective losses. Such passive $\mathcal{PT}$-symmetric systems have been realized in the quantum domain with lossy, single-photons~\cite{LXZ+17,pxdqpt,pxchern,XWZ+19,BWZ+20,ZXB+17,XDW+20,Klauck2019}, ultracold atoms~\cite{Luo19}, and a superconducting transmon~\cite{NAJM19}. These realizations are limited to effective two-dimensional Hamiltonians with second-order EPs, and their quantum information studies are confined to global properties~\cite{XWZ+19}. Here we present experimental quantum simulation of entropy dynamics in a four-dimensional, passive $\mathcal{PT}$-symmetric system with an EP$4$.

\section{Implementing $\mathcal{PT}$-symmetric qudit with an EP$4$}

\begin{figure*}
\centering
\includegraphics[width=\textwidth]{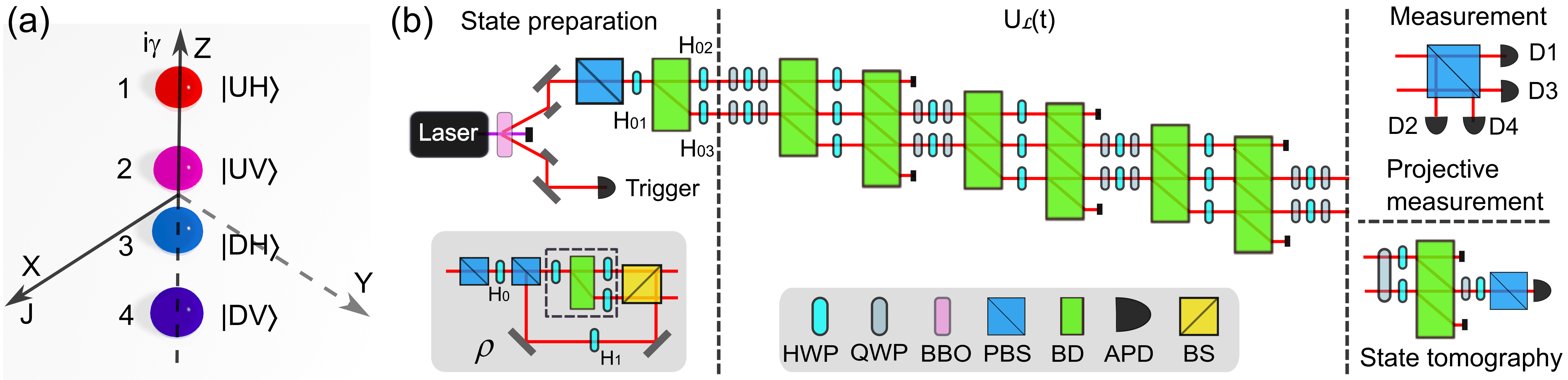}
\caption{Experimental setup. (a) Illustration of a $4$-mode $\mathcal{PT}$-symmetric qudit. (b) Schematic of the optical circuit used for simulating dynamics in the four-mode, passive $\mathcal{PT}$-symmetric system. Heralded single photons are generated via spontaneous parametric down-conversion and prepared in arbitrary qudit states using a polarizing beam splitter (PBS), wave plates with certain setting angles and a beam displacer (BD). The coherent, lossy, non-unitary time evolution is realized by BDs, HWPs, and sandwich-type QWP-HWP-QWP setups, along with single-photon loss. For detection, projective measurements and quantum-state tomography are selected depending on the purpose, either of which is performed, both of which are realized by a PBS and (or) wave plates and a BD. Avalanche photodiodes (APDs) detect the signal and heralding photons.}
\label{fig:1}
\vspace{-3mm}
\end{figure*}

Let us consider an open, four-mode system described by a $4\times 4$ Hamiltonian
\begin{equation}
\label{eq:hpt}
H_\mathcal{PT}=-JS_x+i\gamma S_z,
\end{equation}
where $S_x$ and $S_z$ are spin-$3/2$ representations of the SU(2) group. It can be written in the matrix form as
\begin{equation}
\label{eq:hpt2}
H_\mathcal{PT}=\frac{1}{2}
\begin{pmatrix}
  3i\gamma & -\sqrt{3}J & 0 & 0 \\
-\sqrt{3}J & i\gamma & -2J & 0 \\
0 & -2J & -i\gamma & -\sqrt{3}J\\
0 & 0 & -\sqrt{3}J & -3i\gamma
\end{pmatrix}
\end{equation}
in the computational basis $\{|1\rangle,|2\rangle,|3\rangle,|4\rangle\}$, and represents a $\mathcal{PT}$-symmetric qudit with $d=4$. The Hamiltonian $H_\mathcal{PT}$ commutes with the antilinear $\mathcal{PT}$ operator where the parity operator is $\mathcal{P}=\mathrm{antidiag}(1, 1, 1, 1)$ and a time-reversal operator is given by complex conjugation, $\mathcal{T}=*$. It follows from Eq.~(\ref{eq:hpt2}) that the first two computational modes represent the ``gain sector'' and the last two represent the ``loss sector'' in the system. The four equally spaced eigenvalues of $H_\mathcal{PT}$ are given by $\lambda_k=\{-3/2,-1/2,+1/2,+3/2\}\sqrt{J^2-\gamma^2}$ ($k=1,2,3,4$), which give rise to an EP$4$ at the $\mathcal{PT}$-breaking threshold $\gamma=J$. The advantage of choosing Hamiltonian (\ref{eq:hpt}) is that it can be easily generalized to an arbitrary dimensional system where it still remains analytically solvable and has an EP with the order equal to the system dimension~\cite{HAS+17,EUH+08,QJ2019}. Since $H_\mathcal{PT}$ has a single energy gap $\Delta=\sqrt{J^2-\gamma^2}$, it follows that the $\mathcal{PT}$ symmetric qudit has a sinusoidal dynamics in the $\mathcal{PT}$-symmetry unbroken region ($\gamma<J$), and a monotonic, exponential growth behavior in the $\mathcal{PT}$-broken region ($\gamma>J$).

The coherent, non-unitary time evolution operator for the system is given by $U(t)=\exp(-i H_\mathcal{PT}t)$ where we have set $\hbar=1$. For $\gamma=0$, the system is Hermitian and the fermionic nature of spin-$3/2$ representation is manifest in the anti-periodicity of $U$, i.e., $U(T)=-\mathbb{I}_4$ where $T(0)=2\pi/J$ for $\gamma=0$. In this case, the mode-occupations $P_k(t)=|\langle k |\psi(t)\rangle|^2$ of the four modes obey a shifted mirror symmetry with $P_k(t)=P_{5-k}(t+T/2)$, which indicates a perfect state transfer occurring from mode $k$ to mode $(5-k)$ at $T/2$. Here $|\psi(t)\rangle=U(t)|\psi(0)\rangle$ is the time-evolved state. For $\gamma < J$, the system is in the $\mathcal{PT}$-symmetry unbroken region, the dynamical evolution is anti-periodical with period $T(\gamma)=2\pi/\Delta$. At the EP$4$ ($\gamma= J$), $U(t)$ ceases to be periodic and has an operator norm that grows as $t^6$, reflecting the fourth order of the EP. In the $\mathcal{PT}$-symmetry broken region, the mode occupations grow exponentially with time. However, the quantum information metrics, such as the von Neumann entropy, are defined with respect to the instantaneously normalized state (indicating post-selection that eliminates the quantum jumps~\cite{NAJM19,Klauck2019,QJ2019}). Therefore, at the EP and in the $\mathcal{PT}$-broken region, these quantities reach a steady-state value. These results are applicable to all finite-dimensional representation of the $SU(2)$ group.

The four-dimensional Hamiltonian $H_\mathcal{PT}$ is particularly interesting because it can be viewed as a system of two interacting, non-Hermitian qubits. This mapping is provided by the identities  $2S_x=\sigma_x\otimes\sigma_x+\sigma_y\otimes\sigma_y+\sqrt{3}\mathbb{I}_2\otimes\sigma_x$, $2S_z=\sigma_z\otimes\mathbb{I}_2+\mathbb{I}_2\otimes\sigma_z/2$, and $\mathcal{P}=\sigma_x\otimes\sigma_x$, where $\sigma_k$ ($k=x,y,z$) are the standard Pauli matrices. Using this insight, we investigate the quantum information dynamics in the gain and loss subsystems of the $\mathcal{PT}$-symmetric qudit.

\begin{figure}
\centering
\includegraphics[width=0.5\textwidth]{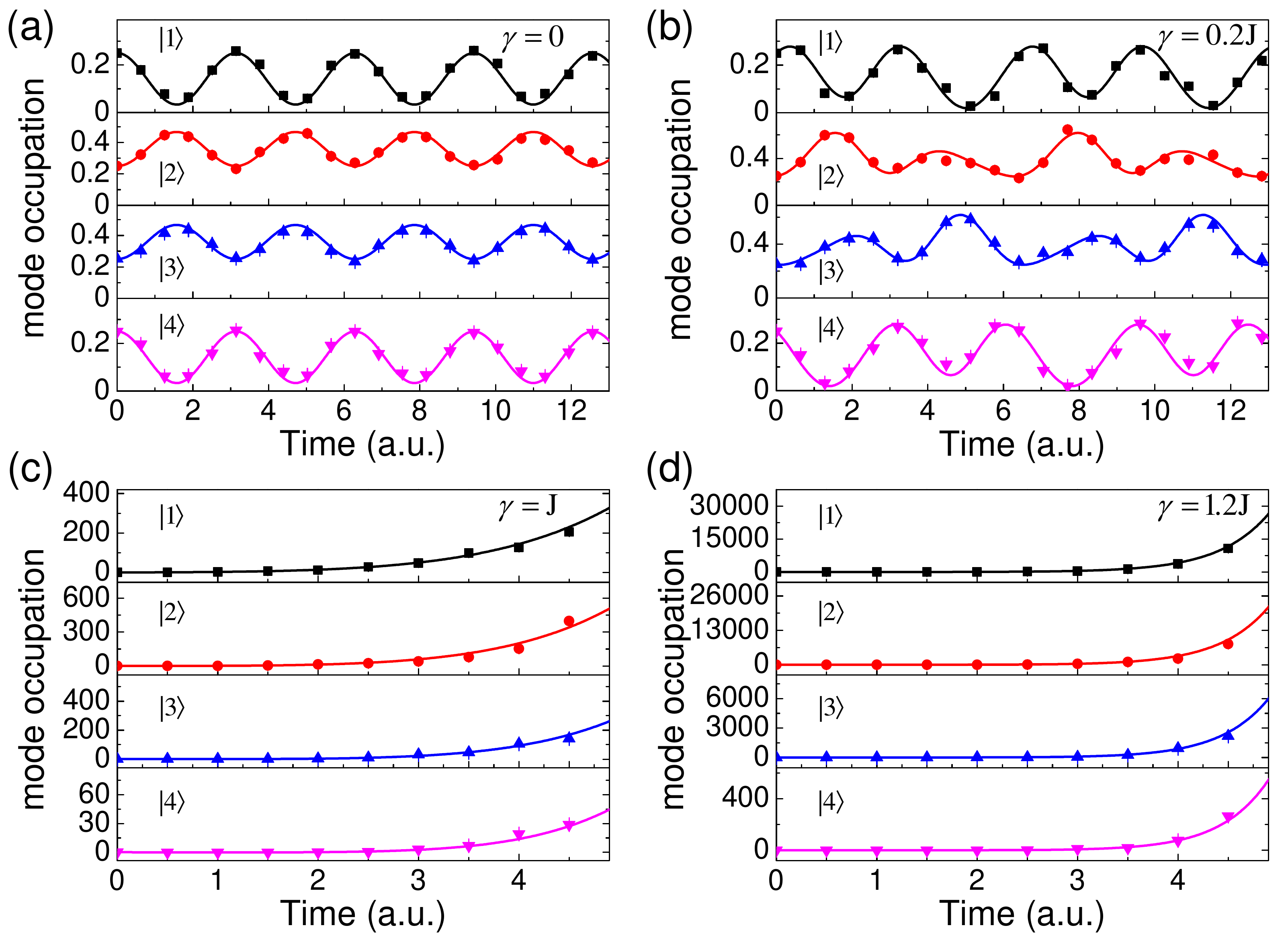}
\caption{Scaled mode occupations of a $\mathcal{PT}$-symmetric system. (a) In the Hermitian limit with $\gamma=0$, mode occupation numbers are periodic with period $T=2\pi/J$, and perfect state transfer occurs from mode $k$ to mode $5-k$ at time $T/2$. (b) In the $\mathcal{PT}$-unbroken phase with $\gamma=0.2J$, the occupation dynamics have period $T(\gamma)$, and the total weights exceed unity indicating a non-trace-preserving time evolution. (c) At $\gamma=J$, the scaled mode occupation grows as $t^6$ with time indicating that the exceptional point is of order four. (d) In the $\mathcal{PT}$-broken phase with $\gamma=1.2J$, the scaled mode occupation grows exponentially with time. Time is measured in units of $J$. Symbols: data; lines: theory. Error bars are due to the statistical uncertainty and obtained based on assuming Poisson statistics. When not shown, error bars are smaller than the symbol size.}
\label{fig:2}
\vspace{-3mm}
\end{figure}
%===============================%

We encode the four modes of the qudit in the spatial and polarization degrees of freedom of a single photon, and label them as $|1\rangle=|UH\rangle,|2\rangle=|UV\rangle,|3\rangle=|DH\rangle,|4\rangle=|DV\rangle$. Here $\{|H\rangle,|V\rangle\}$ are the horizontal and vertical polarizations, and $\{|U\rangle,|D\rangle\}$ denote the upper and lower paths, which undergo gain and loss respectively (Fig.~\ref{fig:1}(a)). As illustrated in Fig.~\ref{fig:1}(b), pairs of single photons are generated via type-I spontaneous parametric down conversion (SPDC) using a non-linear $\beta$-Barium-Borate (BBO) crystal. One photon serves as a trigger and the other signal photon is prepared in an arbitrary qudit state using a polarizing beam splitter (PBS), wave plates with certain setting angles and a beam displacer (BD).

By mapping the $\mathcal{PT}$-symmetric Hamiltonian $H_{\mathcal{PT}}$ into a passive $\mathcal{PT}$-symmetric one with mode-selective losses $H_\mathcal{L}=H_\mathcal{PT}-3i\gamma\mathbb{I}_4/2$, we implement the $4\times4$ lossy, time-evolution operator
\begin{equation}
U_\mathcal{L}(t)=\exp(-iH_{\mathcal{L}}t)
\end{equation}
via a lossy linear optical circuit, which is related to $U(t)$ through $U(t)= U_\mathcal{L}(t)\exp(3\gamma t/2)$~\cite{LXZ+17}. The evolution operator $U_\mathcal{L}(t)$ is realized by BDs, half-wave plates (HWPs), and sandwich-type QWP-HWP-QWP setups, where QWP is an abbreviation for quarter-wave plate.

We experimentally measure and then obtain scaled mode occupations $P_k(t)$ by projecting the time-evolved state $|\psi(t)\rangle$ onto $|k\rangle$. The initial state is chosen to be $|\psi(0)\rangle=(|1\rangle+|2\rangle+|3\rangle+|4\rangle)/2$. The projective measurement and the quantum state tomography on the qudit state are realized by BDs, wave plates and a PBS followed by avalanche photodiodes (APDs). Only coincidences between the heralded and trigger photons are registered. The perfect state transfer for $\gamma=0$ is confirmed by the transfer of occupation from the first mode to the fourth mode (Fig.~\ref{fig:2}(a)). In the $\mathcal{PT}$-unbroken phase with a finite $\gamma=0.2J$, there is no perfect state transfer at time $T(\gamma)/2$ due to the non-unitary dynamics (Fig.~\ref{fig:2}(b)). The measured occupations are, however, periodic in time with a period $T(\gamma)$.

At the EP$4$ with $\gamma=J$, the scaled mode occupation $P_k(t)$ grows algebraically with time as $t^6$ (Fig.~\ref{fig:2}(c)). Such a scaling is dictated by the order of the EP. At $\gamma=J$, the Hamiltonian obeys $H^{4}_\mathcal{PT}(\gamma=J)=0$ and the power-series expansion of $U(t)$ terminates at the third order, giving rise to $t^6$ dependence for the occupation numbers. By projecting the time-evolved state onto $|k\rangle$, we can obtain the occupation at the EP$4$ and its power-law behaviour (Fig.~\ref{fig:2}(c)). In the $\mathcal{PT}$-broken phase, the scaled mode occupation grows exponentially with time as expected (Fig.~\ref{fig:2}(d)). We note that while the simulation time-range is limited to two periods for $\gamma<J$, we restrict to $0\leq t\leq 4.5$ due to the rapid growth of the scaled mode occupation at the EP$4$ and in the broken $\mathcal{PT}$ region.

%==================================%
\begin{figure}
\centering
\includegraphics[width=0.5\textwidth]{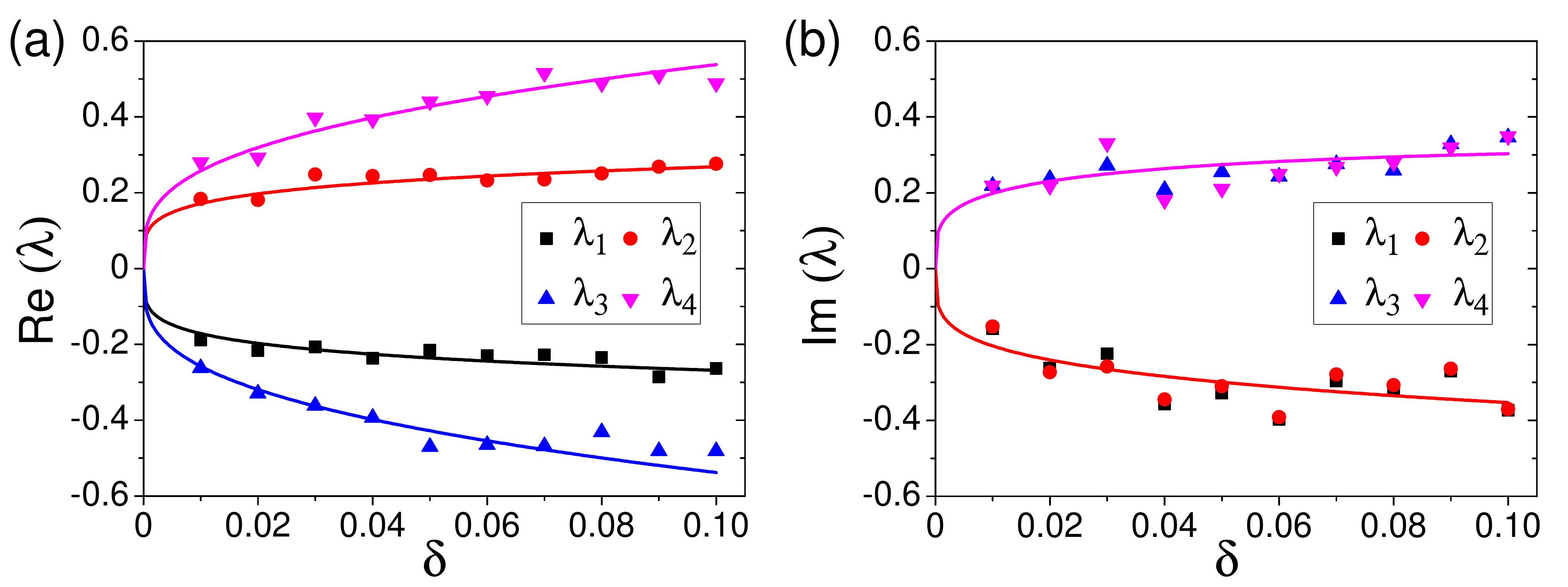}
\caption{Eigenvalues of the perturbed Hamiltonian. The real (a) and imaginary (b) parts of the eigenvalues of the perturbed Hamiltonian $H_\delta=H_\mathcal{PT}(\gamma=J)+iJ\delta|1\rangle\langle 1|$, measured in units of $J$, scale as $\delta^{1/4}$, showing the enhanced sensitivity near the EP$4$. The initial states are eigenstates $|\psi_\delta\rangle$ of $H_\delta$. Experimental errors are calculated via Monte Carlo method; when not shown, error bars are smaller than the symbol size.}
\label{fig:S2}
\vspace{-3mm}
\end{figure}
%==================================%

When the $\mathcal{PT}$-symmetric Hamiltonian is perturbed from the EP$4$ by a small detuning $\delta$, the resulting complex eigenvalues in the vicinity of EP$n$ are given by a Puiseux series in $\delta^{1/n}$~\cite{EUH+08}, indicating enhanced classical sensitivity proportional to the order of the EP~\cite{HAS+17,WSG+17}. In addition to the behavior of the mode occupations at the EP, this serves as a complementary check of the order of the EP. To that end, we experimentally measure the complex eigenvalues of the perturbed Hamiltonian $H_\delta=H_\mathcal{PT}(\gamma=J)+iJ\delta|1\rangle\langle 1|$. Figure~\ref{fig:S2} shows that the real and imaginary parts of the eigenvalues of $H_\delta$ indeed scale as $\delta^{1/4}$, consistent with the EP$4$ that occurs at $\gamma=J$.
%-------------------------------------------------------------------------------------%

\section{Observing information dynamics}

%==========================%
\begin{figure*}
\centering
\includegraphics[width=\textwidth]{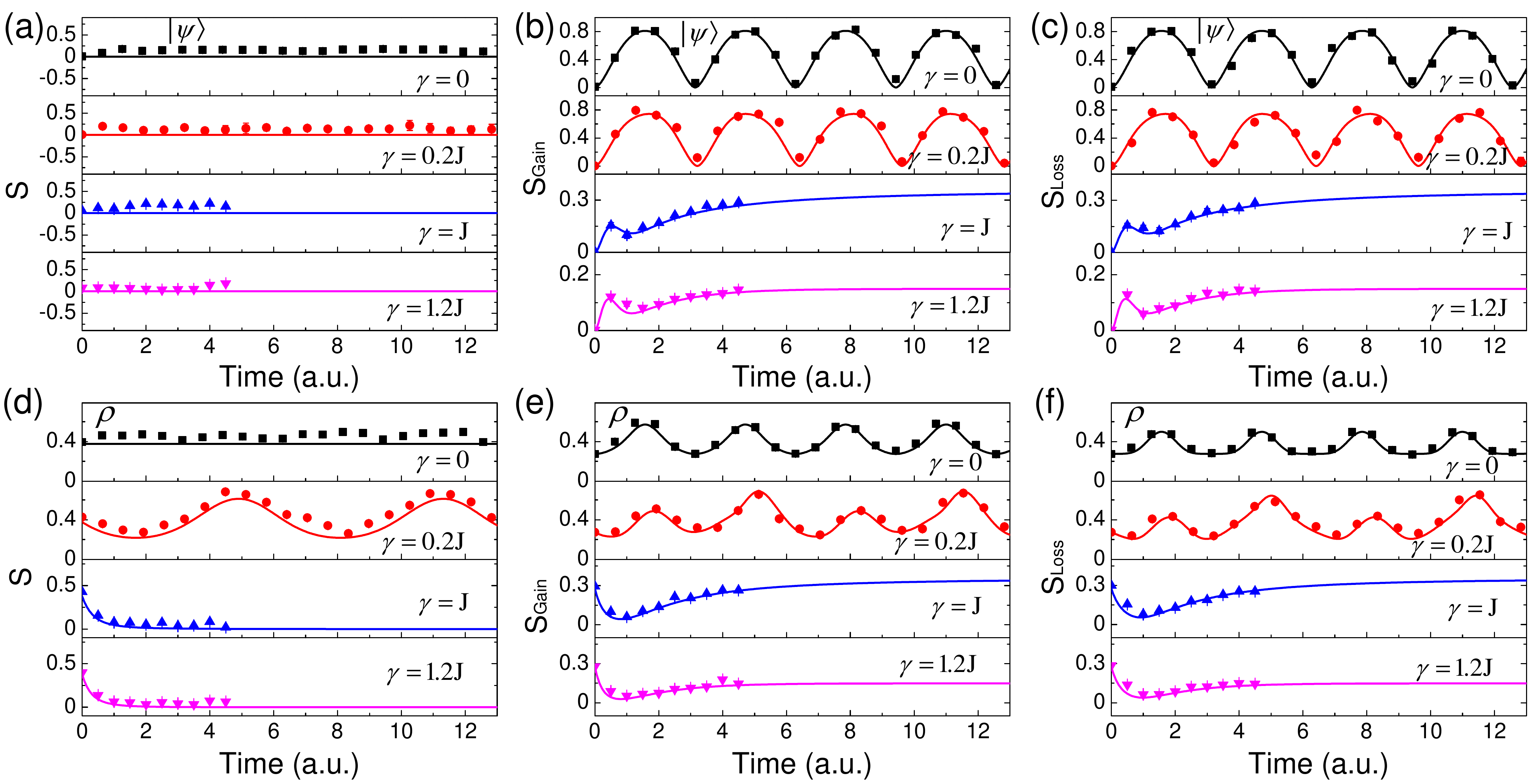}
\caption{Quantum information dynamics of the $\mathcal{PT}$-symmetric qudit. (a) Time-dependent entropy $S(t)$ for a qudit with pure initial state $|\psi(0)\rangle$ remains zero for any non-Hermiticity. (d) With mixed initial state, $S(t)$ is constant when $\gamma=0$, and oscillates in the $\mathcal{PT}$-unbroken phase ($\gamma=0.2J$) with period $T(\gamma)$. At the EP$4$ ($\gamma=J$) and in the broken $\mathcal{PT}$ region, $S(t)$ reaches zero because the system approaches a pure state that is determined by the sole eigenstate at the EP or the mode with maximum amplification. In contrast, the subsystem entropies $S_\mathrm{Gain}(t)$ and $S_\mathrm{Loss}(t)$ show qualitatively similar behavior for pure (b)-(c) and mixed (e)-(f) initial qudit states. In each case, the entropies show oscillatory behavior for $\gamma<J$ and steady-state behavior for $\gamma\geq J$. We restrict to $0\leq t\leq 4.5$ due to the rapid growth of the scaled mode occupation at the EP$4$ and in the broken $\mathcal{PT}$ region. Experimental errors are due to Monte Carlo method; when not shown, error bars are smaller than the symbol size.}
\label{fig:3}
\vspace{-3mm}
\end{figure*}
%============================%

%============================%
\begin{figure*}
\centering
\includegraphics[width=\textwidth]{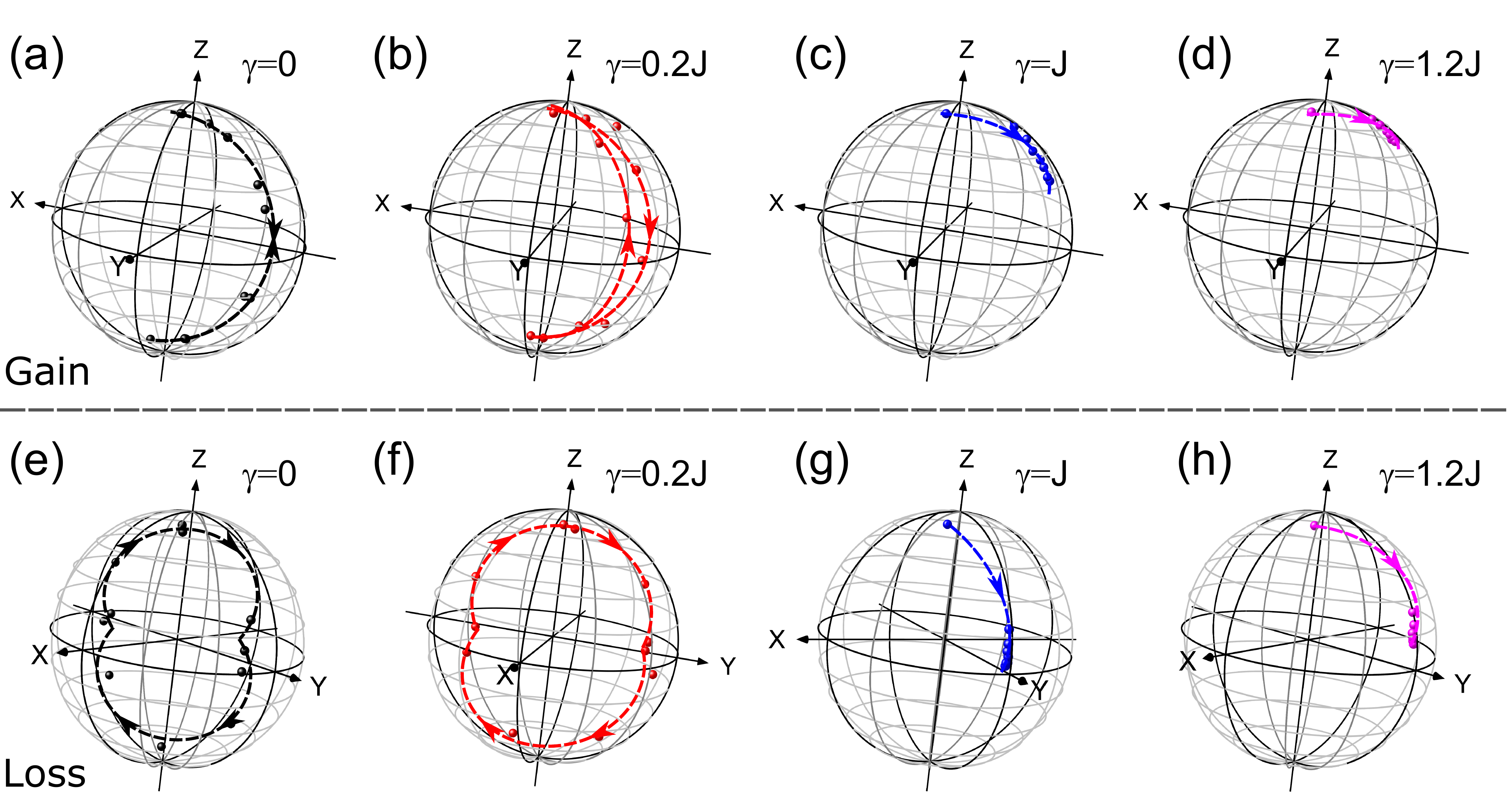}
\caption{Reduced, instantaneously normalized density matrices $\tilde\rho_\mathrm{Gain}(t)$ (a)-(d) and $\tilde\rho_\mathrm{Loss}(t)$ (e)-(h) trace different paths inside the Bloch sphere. Experimental results are represented by colored symbols and their theoretical predictions are represented by dashed curves.}
\label{fig:4}
\vspace{-3mm}
\end{figure*}
%============================%

A crucial aspect of dynamics of a high-dimensional $\mathcal{PT}$-symmetric system is the flow of information among its different parts, and the information retrieval phenomena between the whole system and its environment. To that end, we consider the qudit entropy
\begin{equation}
S(t)=-\text{Tr}\left[\tilde{\rho}(t)\log_2\tilde{\rho}(t)\right],
\end{equation}
where $\tilde{\rho}(t)=\rho(t)/\text{Tr}\left[\rho(t)\right]$ is the instantaneously normalized density matrix and $\rho(t)=U(t)\rho(0)U^\dagger(t)$ is the time-evolved density matrix of the system with a time-dependent trace. The gain- and loss-sector entropies are $S_\mathrm{Gain}(t)$ and $S_\mathrm{Loss}(t)$, respectively. These are obtained from the gain- and loss-sector reduced density matrices $\rho_\mathrm{Gain}(t)=\text{Tr}_{3,4}\left[\rho(t)\right]$ and $\rho_\mathrm{Loss}(t)=\text{Tr}_{1,2}\left[\rho(t)\right]$, respectively.

A full knowledge of the time-dependent density matrix through the quantum-state tomography allows us to experimentally explore the information flow. We focus on the quantum dynamics with the fully symmetric initial state $|\psi(0)\rangle$ (Figs.~\ref{fig:3}(a)-(c)) and a mixed initial state $\rho(0)=0.925|1\rangle\langle1|+0.025\left(|2\rangle\langle2|+|3\rangle\langle3|+|4\rangle\langle4|\right)$ (Figs.~\ref{fig:3}(d)-(f)) in the $\mathcal{PT}$-symmetry unbroken region. Since the qudit undergoes a coherent, non-unitary evolution for any gain-loss strength $\gamma$, a pure state remains a pure state and the entropy of the entire system $S(t)$ remains constant with time (Fig.~\ref{fig:3}(a)). For a mixed initial state, the entropy is constant only in the Hermitian limit, $\gamma=0$. In the $\mathcal{PT}$-symmetry unbroken region, the entropy $S(t)$ shows periodic oscillations. This demonstrates an exchange of quantum information between the $\mathcal{PT}$-symmetric qudit and its environment, and the oscillations observed here may be interpreted as an evidence of information backflow from the environment and a signature of non-Markovianity in the $\mathcal{PT}$-unbroken phase~\cite{XWZ+19}.

At the EP$4$ or in the $\mathcal{PT}$-symmetry broken region, due to the diverging occupation, the normalized density matrix $\tilde\rho(t)$ approaches a pure state, and the total system entropy, therefore, approaches zero~\cite{XWZ+19,KAU17,WLG+19}. In all cases, the experimental simulation results agree well with the theoretical prediction. Importantly, this observed behavior of entropy does not depend on the details of the system, which signifies its universality. In this case, information flows unidirectionally and the dynamics is asymptotically Markovian~\cite{KAU17}.

In a sharp contrast to the results for the entire system, the behavior of subsystem entropies for pure and mixed initial states is qualitatively similar. In either case, the gain-sector entropy $S_\mathrm{Gain}(t)$ and the loss-sector entropy $S_\mathrm{Loss}(t)$ oscillate in the $\mathcal{PT}$-symmetry unbroken region including the Hermitian limit. On the other hand, they reach nonzero steady-state values at EP$4$ and in the broken $\mathcal{PT}$-symmetry region.  It is worth its while to point out that although the gain and loss entropies show qualitatively similar behavior, the trajectories traced out by the instantaneously normalized, reduced density matrices $\tilde\rho_\mathrm{Gain}(t)$ and $\tilde\rho_\mathrm{Loss}(t)$ in the Bloch ball are distinctly different (Fig.~\ref{fig:4}). The trajectory of the gain-sector density matrix is weighted towards the northern hemisphere, representing the largest amplifying mode, whereas the loss-sector density matrix trajectory is less heavily weighted. These differences lead to the slightly different behaviors of $S_\text{Gain}$ and $S_\text{Loss}$.

In this paper, we realize a four-level system dynamics under a non-Hermitian Hamiltonian in either $\mathcal{PT}$-symmetric unbroken, broken or at the exceptional point with single photons and a cascaded interferometric setup. We realize $4\times4$ non-unitary evolution operations with six BDs and use another one for state preparation. Two different measurements---projective measurement and the quantum state tomography of a four-level system---are carried out at the output. In contrast, the setup in~\cite{XWZ+19} is much simpler; a two-level system dynamics under a non-Hermitian Hamiltonian is realized with two BDs, and only a single-qubit state tomography is carried out to reconstruct the final state. Our experimental method to implement a non-unitary, loss time evolution operator is scalable, and therefore can be used to simulate higher-dimensional $\mathcal{PT}$-symmetric systems in the future.

\section{Discussion}
In this section we briefly present the analytical derivation for the entropy of the $\mathcal{PT}$-symmetric system. If we start with a pure state, it remains pure under the coherent, non-unitary evolution that is generated by a $\mathcal{PT}$-symmetric, non-Hermitian Hamiltonian. Therefore, the entropy of such a state continues to remain zero. If the initial state is mixed, i.e., $\rho(0)=\sum_{i}\alpha_i|\upsilon_i\rangle\langle\upsilon_i|$, we can express the orthonormal vectors $|\upsilon_i\rangle=\sum_{k}\beta_{ik}|\zeta_k\rangle$ in terms of the non-orthogonal right eigenvectors $|\zeta_k\rangle$ of $H_\mathcal{PT}$. The initial state, thus, can be rewritten as
\begin{align*}
\rho(0)=\sum_{k,j,i}\alpha_i\beta_{ik}\beta^*_{ij}|\zeta_k\rangle\langle\zeta_j|.
\end{align*}
The final state is then given by
\begin{align*}
\rho(t)=&\sum_{\substack{k,i}}\alpha_i|\beta_{ik}|^2|\zeta_k\rangle\langle\zeta_k|\\
&+\sum_{\substack{k\neq j,i}}\alpha_i\beta_{ik}\beta^*_{ij}e^{-i(\lambda_k-\lambda_j)t}|\zeta_k\rangle\langle\zeta_j|.
\end{align*}
We further express the right eigenvectors of $H_\mathcal{PT}$ in terms of the orthonormal eigenvectors of the instantaneous density matrix $\rho(t)$ as $|\zeta_k\rangle=\sum_l \kappa_{kl}|\varphi_l\rangle$. It allows us to obtain the time-dependent occupation eigenvalues $p_k(t)=\langle\varphi_l|\rho(t)|\varphi_l\rangle$ as
\begin{align*}
p_l=\sum_{k,i,l}\alpha_i|\beta_{ik}|^2|\kappa_{kl}|^2+\sum_{\substack{k\neq j,i,l}}\alpha_i\beta_{ik}\beta^*_{ij}\kappa_{kl}\kappa^*_{jl}e^{-i(\lambda_k-\lambda_j)t}.
\end{align*}

In the Hermitian limit, the eigenvectors of $H_\mathcal{PT}$ are orthonormal, and the time evolution acts as the rotation of coordinates. Therefore the eigenstates $|\varphi_i\rangle$ are unchanged and the entropy remains a constant of motion. In the non-Hermitian case, $\{|\zeta_k\rangle\}$ are not orthonormal, and the time-dependent entropy is then given by
\begin{align*}
S(t)=-\sum_l \tilde{p}_l\log_2\tilde{p_l},
\end{align*}
where the fractional occupations are given by $\tilde{p}_l(t)=p_l(t)/\sum_k p_k(t)$. The entropy of time-evolved state oscillates periodically in $\mathcal{PT}$-symmetric unbroken region. At the EP$4$, $p_l(t)$ grow algebraically with time as $t^6$. By writing $p_l=\lambda_l t^6+\mu_l$ where $\lambda_l$ and $\mu_l$ are constant, it is straightforward to see that the entropy approaches a steady-state value polynomially with time. In contrast, in the $\mathcal{PT}$-broken region, $p_l(t)$ grow exponentially with time, leading to a steady-state value for the entropy that is approaches in an exponential manner.

%\begin{align*}
%&S(t)=-\sum_{l=1}^{4}\frac{p_l(t)}{\sum{p_l(t)}}\log_2 \frac{p_l(t)}{\sum{p_l(t)}}\\&=-\frac{\sum{(\lambda_l t^6+\mu_l)\log_2(\lambda_l t^6+\mu_l)}}{\sum{(\lambda_l t^6+\mu_l)}}+\log_2{\sum{(\lambda_l t^6+\mu_l)}},
%\end{align*}
%which decays polynomially with time at EP$4$.
%
%Similarly to EP$4$, in $\mathcal{PT}$-broken region, $p_l$ grow exponentially with time. We can rewrite $p_l=\lambda_l e^{2t}+\mu_l$, and then the entropy is given by
%\begin{align*}
%&S(t)=-\sum_{l=1}^{4}\frac{p_l(t)}{\sum{p_l(t)}}\log_2 \frac{p_l(t)}{\sum{p_l(t)}}\\& =-\frac{\sum{(\lambda_l e^{2t}+\mu_l)\log_2(\lambda_l e^{2t}+\mu_l)}}{\sum{(\lambda_l e^{2t}+\mu_l)}}+\log_2{\sum{(\lambda_l e^{2t}+\mu_l)}},
%\end{align*}
%which decays exponentially with time in $\mathcal{PT}$-broken region.

\section{Summary}

Higher-dimensional $\mathcal{PT}$ systems, which can be treated as composites of two or more minimal, non-Hermitian, quantum systems, provide a starting point for interacting quantum models with $\mathcal{PT}$-symmetry and EP degeneracies. In this work, we experimentally simulate and observe the quantum information dynamics in a four-dimensional system with EP$4$. We show that the subsystem-entropy behavior for gain or loss subsystems can be either qualitatively different from or similar to the dynamics for the total entropy of the four-dimensional system. Our work is the first experimental demonstration of critical phenomena in four-dimensional $\mathcal{PT}$-symmetric quantum dynamics, and shows the versatility of the single-photon interferometric network platform for simulating interacting, non-Hermitian, quantum systems.

\acknowledgements The authors thank support from the National Natural
Science Foundation of China (Grant Nos. 11674056, 11974331 and U1930402), the National Natural
Science Foundation of Jiangsu Province (Grant No. BK20190577), the Fundamental Research Funds for the Central Universities (JUSRP11947), the National Key R\&D Program (Grant Nos. 2016YFA0301700 and 2017YFA0304100) and NSF DMR-1054020.

\bibliographystyle{plain}
\bibliographystyle{apsrev4-1}
\bibliography{EPentropy.bib}

\end{document}